\documentclass[twocolumn,showpacs,aps,prl,superscriptaddress]{revtex4}

\usepackage{graphicx}
\usepackage{dcolumn}
\usepackage{amsmath}
\usepackage{epsfig}

\RequirePackage{xspace}

\usepackage{relsize}
\def\babar{\mbox{\slshape B\kern-0.1em{\smaller A}\kern-0.1em
    B\kern-0.1em{\smaller A\kern-0.2em R}}}

\def\pip   {\ensuremath{\pi^+}\xspace}
\def\pim   {\ensuremath{\pi^-}\xspace}

\def\Kbar  {\kern 0.2em\overline{\kern -0.2em K}{}\xspace}

\def\Kz    {\ensuremath{K^0}\xspace}
\def\Kzb   {\ensuremath{\Kbar^0}\xspace}
\def\KzKzb {\ensuremath{\Kz \kern -0.16em \Kzb}\xspace}
\def\Kp    {\ensuremath{K^+}\xspace}
\def\Km    {\ensuremath{K^-}\xspace}

\def\KpKm  {\ensuremath{\Kp \kern -0.16em \Km}\xspace}

\def\Dbar    {\kern 0.2em\overline{\kern -0.2em D}{}\xspace}

\def\Dz      {\ensuremath{D^0}\xspace}
\def\Dzb     {\ensuremath{\Dbar^0}\xspace}
\def\DzDzb   {\ensuremath{\Dz {\kern -0.16em \Dzb}}\xspace}
\def\Dp      {\ensuremath{D^+}\xspace}
\def\Dm      {\ensuremath{D^-}\xspace}

\def\DpDm    {\ensuremath{\Dp {\kern -0.16em \Dm}}\xspace}

\def\B       {\ensuremath{B}\xspace}
\def\Bbar    {\kern 0.18em\overline{\kern -0.18em B}{}\xspace}

\def\BB      {\ensuremath{B\Bbar}\xspace} 
\def\Bz      {\ensuremath{B^0}\xspace}
\def\Bzb     {\ensuremath{\Bbar^0}\xspace}
\def\BzBzb   {\ensuremath{\Bz {\kern -0.16em \Bzb}}\xspace}
\def\Bu      {\ensuremath{B^+}\xspace}
\def\Bub     {\ensuremath{B^-}\xspace}

\def\Bm      {\ensuremath{\Bub}\xspace}

\def\BpBm    {\ensuremath{\Bu {\kern -0.16em \Bub}}\xspace}

\def\BorBbar    {\kern 0.18em\optbar{\kern -0.18em B}{}\xspace}
\def\DorDbar    {\kern 0.18em\optbar{\kern -0.18em D}{}\xspace}
\def\KorKbar    {\kern 0.18em\optbar{\kern -0.18em K}{}\xspace}

\def\jpsi     {\ensuremath{{J\mskip -3mu/\mskip -2mu\psi\mskip 2mu}}\xspace}
\def\psitwos  {\ensuremath{\psi{(2S)}}\xspace}

\mathchardef\Upsilon="7107
\def\Y#1S{\ensuremath{\Upsilon{(#1S)}}\xspace}

\def\FourS {\Y4S}

\mathchardef\Deltares="7101
\mathchardef\Xi="7104
\mathchardef\Lambda="7103
\mathchardef\Sigma="7106
\mathchardef\Omega="710A

\def\Deltabar{\kern 0.25em\overline{\kern -0.25em \Deltares}{}\xspace}
\def\Lbar{\kern 0.2em\overline{\kern -0.2em\Lambda\kern 0.05em}\kern-0.05em{}\xspace}
\def\Sigbar{\kern 0.2em\overline{\kern -0.2em \Sigma}{}\xspace}
\def\Xibar{\kern 0.2em\overline{\kern -0.2em \Xi}{}\xspace}
\def\Obar{\kern 0.2em\overline{\kern -0.2em \Omega}{}\xspace}
\def\Nbar{\kern 0.2em\overline{\kern -0.2em N}{}\xspace}
\def\Xb{\kern 0.2em\overline{\kern -0.2em X}{}\xspace}

\def\BR         {{\ensuremath{\cal B}\xspace}}

\def\mes        {\mbox{$m_{\rm ES}$}\xspace}

\def\DeltaE     {\mbox{$\Delta E$}\xspace}

\newcommand{\tev}{\ensuremath{\mathrm{\,Te\kern -0.1em V}}\xspace}
\newcommand{\gev}{\ensuremath{\mathrm{\,Ge\kern -0.1em V}}\xspace}
\newcommand{\mev}{\ensuremath{\mathrm{\,Me\kern -0.1em V}}\xspace}
\newcommand{\kev}{\ensuremath{\mathrm{\,ke\kern -0.1em V}}\xspace}
\newcommand{\ev}{\ensuremath{\mathrm{\,e\kern -0.1em V}}\xspace}
\newcommand{\gevc}{\ensuremath{{\mathrm{\,Ge\kern -0.1em V\!/}c}}\xspace}
\newcommand{\mevc}{\ensuremath{{\mathrm{\,Me\kern -0.1em V\!/}c}}\xspace}
\newcommand{\gevcc}{\ensuremath{{\mathrm{\,Ge\kern -0.1em V\!/}c^2}}\xspace}
\newcommand{\mevcc}{\ensuremath{{\mathrm{\,Me\kern -0.1em V\!/}c^2}}\xspace}

\def\mus  {\ensuremath{\rm \,\mus}\xspace}

\def\mus        {\ensuremath{\,\mu{\rm s}}\xspace}

\def\to                 {\ensuremath{\rightarrow}\xspace}

\def\pep2{PEP-II}

\def\gsim{{~\raise.15em\hbox{$>$}\kern-.85em
          \lower.35em\hbox{$\sim$}~}\xspace}
\def\lsim{{~\raise.15em\hbox{$<$}\kern-.85em
          \lower.35em\hbox{$\sim$}~}\xspace}

\xspace

\newcommand{\jprlBase}       {Phys.\ Rev.\ Lett.\xspace}
\newcommand{\jprBase}        {Phys.\ Rev.\xspace}

\newcommand{\nimBaseC}       {Nucl.\ Instr.\ and Methods\xspace}

\newcommand{\zpBase}         {Z.\ Phys.\xspace}

\newcommand{\nim}       [1]  {\nimBaseC~{\bf #1}}

\newcommand{\jprl}      [1]  {\jprlBase\ {\bf #1}}

\newcommand{\jprd}      [1]  {\jprBase\ D~{\bf #1}}

\newcommand{\zp}        [1]  {\zpBase\ {\bf #1}}

\def\jetset74   {\mbox{\tt Jetset \hspace{-0.5em}7.\hspace{-0.2em}4}\xspace}

\newcommand{\nbb}{117}

\newcommand{\btopsik}{\mbox{\ensuremath{\Bm\to\psitwos K^-}}}
\newcommand{\btojpsikpp}{\mbox{\ensuremath{\Bm\to\jpsi K^-\pip\pim}}}
\newcommand{\rmeas} {\ensuremath{1.70 \pm 0.10 (stat.) \pm 0.09(syst.)}}
\newcommand{\brjpsikppval}{116}
\newcommand{\brjpsikpperr}{\ensuremath{ 7 (stat.) \pm 9(syst.)}}
\newcommand{\brjpsikpp}{\ensuremath{(\brjpsikppval \pm \brjpsikpperr)\times~10^{-5}}}
\newcommand{\btoxk}{\ensuremath{\Bm\to\it{X}\rm{(3872)}\it{K^-}}}
\newcommand{\brxk}{\ensuremath{ \rm (1.28\pm 0.41)\times 10^{-5}}}
\newcommand{\btohck}{\ensuremath{\Bm\to h_c K^-}}
\newcommand{\brhck}{\ensuremath{ 3.4 \times 10^{-6} }}
\newcommand{\btojdp}{\ensuremath{\Bm\to\jpsi \Dz\pim}}
\newcommand{\brjdp}{\ensuremath{ 5.2 \times 10^{-5} }}

\newcommand{\brf}{branching fraction}
\def\mjpp       {\ensuremath{m_{J/\psi\pi\pi}}}
\def\mkp       {\ensuremath{m_{\Km\pip}}}
\def\mpipi     {\ensuremath{m_{\pip\pim}}}
\def\myprl  #1 #2 #3 {\jprl{#1},\ #2 (#3)}
\def\myprd  #1 #2 #3 {\jprd{#1},\ #2 (#3)}
\def\mynim  #1 #2 #3 {\nim{#1},\ #2 (#3)}

\newcommand{\BABARPubYear}    {04}
\newcommand{\BABARPubNumber}  {011}

\newcommand{\SLACPubNumber} {10475}

\def\figurebox#1#2#3{
    \def\arg{#3}
    \ifx\arg\empty
    {\hfill\vbox{\hsize#2\hrule\hbox to #2{\vrule\hfill\vbox to #1{\hsize#2\vfill}\vrule}\hrule}\hfill}
    \else
    {\hfill\epsfbox{#3}\hfill}
    \fi}

\begin{document}

\begin{flushleft}
\babar-PUB-\BABARPubYear/\BABARPubNumber\\
SLAC-PUB-\SLACPubNumber\\[20mm]
\end{flushleft} 


\title{
{\large \boldmath
Study of the \btojpsikpp\ Decay and Measurement of the \btoxk\ Branching Fraction.
}
}

\author{B.~Aubert}
\author{R.~Barate}
\author{D.~Boutigny}
\author{F.~Couderc}
\author{J.-M.~Gaillard}
\author{A.~Hicheur}
\author{Y.~Karyotakis}
\author{J.~P.~Lees}
\author{V.~Tisserand}
\author{A.~Zghiche}
\affiliation{Laboratoire de Physique des Particules, F-74941 Annecy-le-Vieux, France }
% ...
\collaboration{The \babar\ Collaboration}
\noaffiliation

\date{\today}

\begin{abstract}

We study the decay 
$B^- \rightarrow J/\psi K^- \pi^+ \pi^-$ using
117 million $B\bar B$ events collected at the 
$Y(4S)$ resonance with the BaBar detector 
at the PEP-II $e^+ e^-$ asymmetric-energy storage ring. 
We measure the branching fractions
$\cal B$ $(B^- \rightarrow J/\psi K^- \pi^+ \pi^-)=
(116 \pm 7 (stat.) \pm 9(syst.))\times~10^{-5}$ 
and 
\hbox{ $\cal B$ $(B^- \rightarrow \it{X}\rm{(3872)}\it{K^-})\times $
$\cal B$ $(X \rm{(3872)}\rightarrow J/\psi \pi^+ \pi^-)
=\rm (1.28\pm 0.41)\times 10^{-5}$}
and find the mass of the $X$(3872) to be \hbox{$3873.4 \pm 1.4 \rm MeV/c^2$}. 
We search for the $h_c$ narrow state
in the decay 
$\B^- \rightarrow h_c K^-$, $h_c\rightarrow J/\psi \pi^+\pi^-$ and for the decay
$B^-\rightarrow J/\psi D^0\pi^-$, with $D^0 \rightarrow K^-\pi^+$. 
We set the 90\% C.L. limits
$\cal B$ $(\B^- \rightarrow h_c K^-)$ $\times $
$\cal B$ $(h_c\rightarrow J/\psi \pi^+\pi^-) <3.4 \times 10^{-6}$ 
and  $\cal B$ $(B^-\rightarrow J/\psi D^0\pi^- ) <5.2 \times 10^{-5}$.

\end{abstract}

\pacs{13.25.Hw, 12.15.Hh, 11.30.Er}

\maketitle

The study of $B$ decays to final states containing 
charmonium and strange mesons is especially suited to 
the search for new charmonium states and for intrinsic charm.
In particular, the decay \btojpsikpp~\cite{cc}
can occur via the production of charmonium states decaying into
$\jpsi\pip\pim$ or possibly via
\btojdp, with $\hbox{\Dz\to\Km\pip}$.
Recently the Belle~\cite{belle} and CDF~\cite{cdf} collaborations have
observed a new state, the $X$(3872),
decaying into $\jpsi\pip\pim$. This state is a charmonium
candidate, the $1^{3}D_{2}$ or $1^{3}D_{3}$~\cite{quigg}, 
with $J^{PC}=2^{--}$ or $J^{PC}=3^{--}$, 
or even  possibly a molecule of charmed $D$ and $D^*$ mesons~\cite{pakvasa}. 
In this Letter, using \nbb million \FourS\ decays into \BB pairs,
we confirm the observation of the $X$(3872) and search for the unconfirmed
charmonium $1P_1$ state $h_c$(3526)~\cite{hc}.
In addition, we study final states involving $D$ mesons to test
models developed to explain the excess of 
low momentum \jpsi mesons  in inclusive 
$B$ decays~\cite{jpsiincl}. 

The data were collected  at the PEP-II asymmetric-energy $e^{+}e^{-}$ 
B-factory with the
 $\babar$ detector, which is fully described elsewhere~\cite{babar-det}.
The detector includes a silicon vertex tracker and a 
drift chamber in a 1.5-T solenoidal magnetic field, which
detect charged particles and measure their momentum and
energy loss. Photons, electrons, and neutral hadrons are detected in a
CsI(Tl)-crystal electromagnetic calorimeter. 
A ring-imaging Cherenkov detector is used
for particle identification. 
Penetrating muons and neutral hadrons are identified  
by resistive-plate chambers in the steel of the flux return.
We use a Monte Carlo simulation of the \babar\ detector based on
GEANT4~\cite{geant}  to validate the analysis procedure and to estimate
efficiency corrections.

The event reconstruction and selection follow closely those
described in an earlier
paper~\cite{babar-charmonium}. The present analysis has been optimized to 
maximize the sensitivity to \btojpsikpp\ decays. 
We reconstruct \hbox{$\jpsi\to e^+e^-$} candidates from pairs of tracks
selected with criteria that are 98\% (7\%) efficient for electrons
(pions).
To account for energy losses, we combine the electron pairs with 
bremsstrahlung-photon candidates and use an 
asymmetric mass window, $2.95<m_{ee(\gamma)}<3.14 \gevcc$.
We reconstruct $\jpsi\to \mu^+\mu^-$ candidates from pairs of 
tracks selected with criteria that are 77\% (8\%) efficient
for muons (pions), satisfying $3.06<m_{\mu\mu}<3.14 \gevcc$. 
The nominal \jpsi\ mass ~\cite{PDG} is imposed as a constraint on \jpsi
candidates, thereby improving the resolution on the $B$ mass and on any
charmonium states in its decay.
Kaons are identified using criteria that have
an efficiency of 97\%, with a 15\% pion-misidentification rate.
$B$-meson candidates are formed by combining a \jpsi\ candidate 
with a kaon candidate and two additional oppositely charged tracks.
To suppress further the background from light-quark production, which is 
characterized by back-to-back jets, 
the angle $\theta_T$ between the thrust axes of the reconstructed 
$B$ candidate and the rest of the event in the center-of-mass system is 
required to satisfy $|\cos\theta_T|<0.8(0.9)$ for $\jpsi\to e^+e^-$ 
($\jpsi\to \mu^+\mu^-$) candidates.

\begin{figure}[!tbp]
\begin{center}
\epsfig{figure=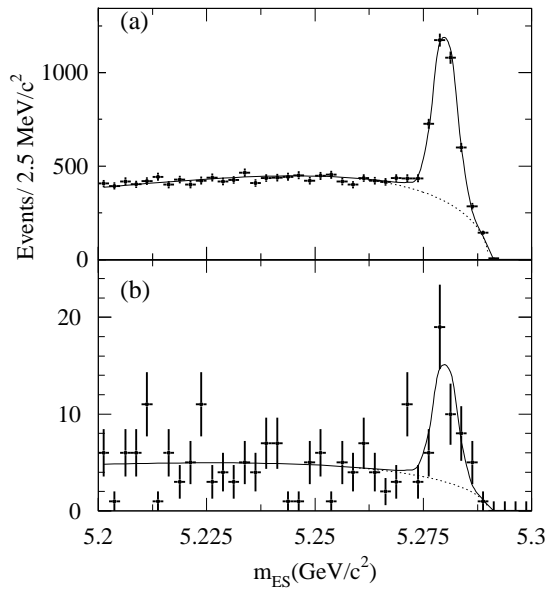,width=1.\linewidth}
\end{center}
\caption{Distribution of \mes\ for (a) \btojpsikpp\ candidates,
and (b) events in the $X$(3872) region, $3862<\mjpp<3882$\mevcc. The solid curves represent
the binned likelihood fits described in the text; the
combinatorial components are indicated by the dashed curves.}
\label{fig:mes}
\end{figure}

Signal and combinatorial 
background are discriminated using two kinematic variables:
the beam-energy--substituted mass,
$\mes~\equiv~\sqrt{(\sqrt{s}/2)^{2} - {p_B^*}^2}$,
and the difference of the $B$ candidate's measured energy from the beam energy, 
$\DeltaE \equiv E_{B}^* - (\sqrt{s}/2)$.
Here $E_{B}^*$ ($p_B^*$) is the energy (momentum) of the \B\ candidate
in the center-of-mass frame and $\sqrt{s}$ is the
total center-of-mass energy.
The signal region is defined to be $|\DeltaE| <3\sigma$, where the 
resolution $\sigma$, determined with data, is 12\mev. 
A binned likelihood fit to the \mes distribution (Fig.~\ref{fig:mes}(a)) is used to 
separate the signal, taken as a Gaussian distribution
with a fitted width of about 2.5\mevcc, plus
a small tail to account for energy losses~\cite{cry},
from the combinatorial background distributed as an ARGUS
threshold function~\cite{Argus}. We have checked with Monte Carlo 
simulation that there is no significant background from $B$ decays that 
has the same \mes\ distribution as the signal.

\begin{figure}[!tbp]
\begin{center}
\epsfig{figure=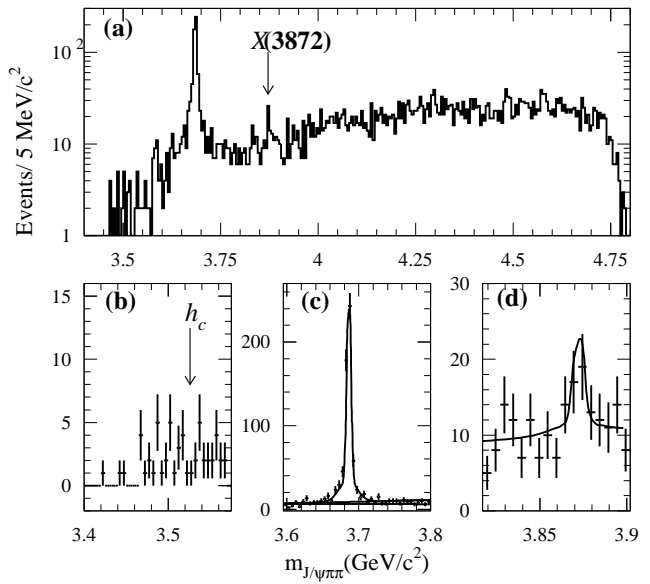,width=1.\linewidth}
\end{center}
\caption{Distribution of \mjpp\ (a) in the entire range, (b)
in the  $h_c$ region, (c) at the \psitwos, and (d)
in the region of the $X$(3872) with the projection of
the unbinned likelihood fit superimposed. The requirement $\mes > 5.27 \gevcc$ is applied.}
\label{fig:xmass}
\end{figure}

To reduce systematic uncertainties, we measure 
\begin{eqnarray}
\label{eq:psi2s}
R &=& \frac{\BR(\btojpsikpp)}{\BR(B^-\rightarrow \psitwos\ K^-)} 
\\ \nonumber &=& \frac{N_{events}}{N_{\psitwos}} \frac{
\epsilon_{\psitwos}}{\epsilon} \times \BR(\psitwos\rightarrow \jpsi \pip \pim),
\end{eqnarray}
where $N_{events}=2540\pm 72$ is the number of \btojpsikpp\ signal events 
extracted from the fit to the \mes\ distribution.
The number of $\psitwos$ events,
$N_{\psitwos}=556 \pm 30$, is obtained 
by fitting \mjpp\ distribution, after subratcting combinatorial background,
with two Gaussian distributions representing the $\psitwos$ signal 
and a flat distribution representing the remaining background.
(Fig.~\ref{fig:xmass}(c) shows the corresponding unsubtracted distribution).
Throughout this Letter the distributions after combinatorial-background
subtraction
are obtained by fitting the \mes\ distribution of the events within each 
bin of the variable of interest (\mjpp\ in this case). 
The binned $\chi^2$ fit gives a resolution 
on \mjpp\ of $3.1 \pm 0.2\mevcc$ for the core Gaussian containing 
70\% of the events and $12\pm3$\mevcc for the broader Gaussian.
The total \btojpsikpp and the \psitwos selection efficiencies, $\epsilon$ 
and $\epsilon_{\psitwos}$, are extracted from Monte Carlo simulation: we obtain 
$\epsilon_{\psitwos} / \epsilon =1.17 \pm 0.03$. We use
\mbox{$\BR(\psitwos\rightarrow \jpsi \pip \pim)=(31.8\pm1.0)\%$~\cite{PDG}.} 

We estimate the systematic error due to the choice of the 
signal \mes\ shape function by replacing it with a 
simple Gaussian. We estimate the uncertainty on the fit to 
the \mjpp\ distribution by using the signal resolution function as 
measured on Monte Carlo and by varying
the background shape. Including all these errors, we
measure $R=\rmeas$ which, combined with 
$\BR(B^-\rightarrow \psitwos\ K^-)=(6.8 \pm 0.4)\times 10^{-4}$~\cite{PDG}, 
yields
\begin{eqnarray}
\BR(\btojpsikpp)&=& \\
  (\brjpsikppval&\pm&\brjpsikpperr)\times10^{-5}.\nonumber
\end{eqnarray}

To investigate the possible presence of narrow charmonium states decaying 
to $\jpsi\pim\pip$, we have studied the distribution in \mjpp\
(Fig.~\ref{fig:xmass}(a)).
We observe an excess in the region of the $X$(3872)
(Fig.~\ref{fig:xmass}(d)), but do not 
find any excess in the $h_c$ region (Fig.~\ref{fig:xmass}(b)).
The mass of the $X$(3872) state is extracted from an unbinned maximum 
likelihood fit to the two-dimensional distribution in \mes\ and \mjpp\ .
The probability density function (PDF) is taken to be the sum of four terms.
The first three describe \btojpsikpp\ decays that peak when \mes is
the mass of the $B$-meson. 
The PDF of these three terms contains a Gaussian function in \mes
times a function of \mjpp\ that describes:
1) non-resonant events, distributed as a first order polynomial; 
2) \psitwos\ candidates, distributed as a double-Gaussian resolution 
function around a mean value that is allowed to float;
and 3) $X$(3872) candidates, with the same resolution function
as the \psitwos but with a mass that floats relative to
the \psitwos mass. The measurement of mass difference allows us to 
neglect systematic errors on the absolute mass scale. 
The fourth term of the PDF describes the combinatorial background, 
distributed as an ARGUS threshold function in \mes\ and as a first order 
polynomial in \mjpp.
From the \psitwos\ mass value, 
$m_{\psitwos}=3685.96\pm0.09\mevcc$~\cite{PDG}, we find
$m_{X(3872)}= 3873.4 \pm 1.4 \mevcc $, consistent with the previous
measurements by Belle~\cite{belle} and CDF~\cite{cdf}.

The measurement of the \brf\ 
\hbox{$\BR ( B^-\rightarrow X(3872) K^-) \times \BR (X(3872) \rightarrow \jpsi \pip \pim)$} 
is performed with a counting technique.
We select events in a $\pm$10\mevcc\ window around \mjpp=3872\mevcc, 
and find the number of events with $\mes>5.27\gevcc$ to be
$N_{data}=63$.
We estimate the number of these events due to combinatorial 
background ($N_{comb}=22.0\pm 4.3$) from a 
fit to the \mes\ distribution (Fig.~\ref{fig:mes}(b)). The number of 
events with the same final state \btojpsikpp,
but not belonging to the $X$(3872) signal, is estimated
to be $N_{peak}=10.5 \pm 3.2$ from a fit to the \mes\ distribution in the 
symmetric sideband $15 < |\mjpp-3872| < 45 \mevcc$. 
The resulting number of signal events is $30.5$ which agrees within the errors 
with the number of signal events, $25.4 \pm 8.7$, obtained from the fit to the 
$X(3872)$ in Figure ~\ref{fig:xmass}(d).
The branching fractions are determined using a frequentist confidence 
level~\cite{PDGstat}.
This technique treats properly the small number of events and includes the 
systematic errors directly in the computation of confidence intervals 
or limits. The confidence level, $\alpha$, 
a function of 
\hbox{$\BR ( B^-\rightarrow X(3872) K^-) \times \BR (X(3872) \rightarrow \jpsi \pip \pim)$} 
is computed as the fraction of times that a
random number generated according to a Poisson distribution with a 
mean value of
\begin{eqnarray}
\label{eq:freq}
\mu &=& N_{bkg} + N_{\psitwos}\epsilon_w
\\ \nonumber &&\times\frac{\BR(\btoxk) \BR(X(\rm{3872})\rightarrow \jpsi \pip \pim)}{\BR(\btopsik) \BR(\psitwos\rightarrow \jpsi \pi^+ \pi^-)}
\end{eqnarray}
exceeds the observed data. 
For a given value~of
\hbox{$\BR ( B^-\rightarrow X(3872) K^-) \times \BR (X(3872) \rightarrow \jpsi \pip \pim)$} 
the variables $N_{bkg}$, $N_{\psitwos}$, 
$\BR( B^- \rightarrow \psitwos K^-)$, 
and $\BR(\hbox{\ensuremath{\psitwos \rightarrow \jpsi \pip \pim}})$
are randomly generated to determine a value of $\mu$, which is 
then used in a Poisson distribution to generate a 
new value of the number of detected events.
The generation is repeated many times and the fraction of times the random 
number exceeds $N_{data}=63$ yields the value of $\alpha$. The 
variables $N_{bkg}$, $N_{\psitwos}$, 
$\BR( B^-\rightarrow\psitwos K^-) $ , 
and $\BR( \psitwos \rightarrow \jpsi \pip \pim)$, 
are generated according to Gaussian distributions. 
The mean of $N_{\psitwos}$ is $556$ and 
$\sigma = 30$. The mean of $N_{bkg}$ is 
$N_{comb}$ $+N_{peak}=32.5$ and $\sigma = 5.9$, which includes a
systematic error on $N_{peak}$ calculated by varying the boundaries
of the sideband.
We use published values~\cite{PDG} for the remaining branching fractions
and their errors, assumed to be Gaussian.
Finally, $\epsilon_w = (92 \pm 1) \%$ is the fraction of events that fall
in the \mjpp\ window, from applying the same mass window cut 
to the \psitwos\ and assuming the same efficiency. 
From the values of \BR(\btoxk) at which $\hbox{\ensuremath{\alpha=16\%}}$ and 
84\% we measure
\begin{eqnarray}\nonumber
\BR(\btoxk)&\times& \BR(X(3872)\rightarrow \jpsi \pi^+ \pi^-)=
\\ && \brxk .
\end{eqnarray}
The probability that the observed events are a background fluctuation in the 
considered mass window is $5.4\times10^{-4}$,
corresponding to 3.5 Gaussian standard deviations.
 As a check, we performed the same measurement on the 
$\hbox{\ensuremath{\jpsi\to e^+e^-}}$ 
and $\hbox{\ensuremath{\jpsi\to\mu^+\mu^-}}$ samples separately, 
obtaining 
$\hbox{\ensuremath{\BR(\btoxk) \times \BR(X(\rm{3872})\rightarrow \jpsi \pip \pim) =}}$
$(1.94\pm0.62)\times 10^{-5}$ and  
$(0.52\pm0.46)\times 10^{-5}$ respectively, consistent within 1.8 standard deviations.

\begin{figure}[!tbp]
\begin{center}
\epsfig{figure=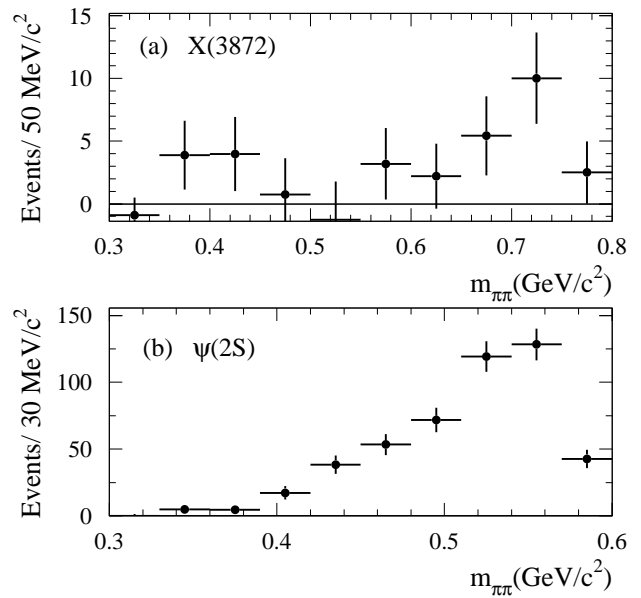,width=1.0\linewidth}
\end{center}
\caption{Distribution of \mpipi (a) at the $X$(3872)
and (b) at the \psitwos, after subtraction of combinatorial and
peaking background.}
\label{fig:mpipi}
\end{figure}

The decay of a charmonium state into $\rho$\jpsi is a strongly suppressed
isospin-violating process. In order to investigate the nature of the
$X$(3872) state, we plot the invariant mass of the \pip\pim system
in both the $X$(3872) and the \psitwos region (Fig.~\ref{fig:mpipi}).
In the \psitwos case, the events are concentrated near the kinematic 
limit. Such behavior is not excluded for the $X$(3872), but the statistics are 
too small to allow a clear conclusion.  Measuring both the \mpipi and 
angular distributions with significantly greater statistics would provide 
important information on the nature of the $X$(3872).

The search for the $h_c$ is performed with the same
frequentist technique
in a $\pm$10\mevcc\ mass window centered on $\mjpp=3526\mevcc$~\cite{hc}.
With $N_{data}=9$, $N_{comb}=6.9\pm 3.5$, $N_{peak}=0.6\pm 1.5$, and
assuming the same efficiency $\epsilon_w = (92 \pm 1) \%$, 
we set a 90\% C.L. limit 
$\hbox{\ensuremath{\BR(\btohck)\times \BR(h_c\rightarrow \jpsi \pip\pim)<\brhck}}$.
The probability that we would see a signal as large as the one observed
from background fluctuations alone is 39\%.

\begin{figure}[!tbp]
\begin{center}
\epsfig{figure=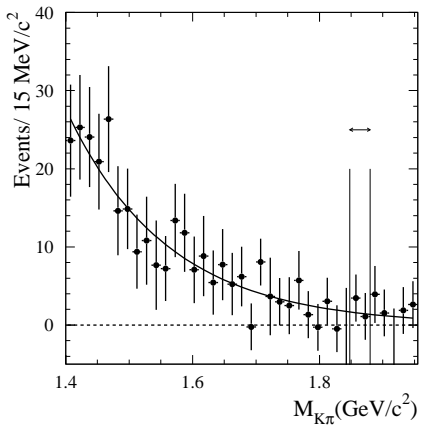,width=0.7\linewidth}
\end{center}
\caption{Distribution of \mkp\ in events \btojpsikpp, with combinatorial 
background removed. Overlaid is an exponential fit. The arrow indicates 
the 3$\sigma$ region expected for $\Dz\to\Km\pip$.}
\label{fig:mkp}
\end{figure}

Finally, we search for \btojdp\ decays with $\Dz\to\Km\pip$.
The decay $\Dz\to\Km\pip$ would have an r.m.s. width of 5.4\mevcc\ 
in \mkp\, as determined from Monte Carlo. We study
this distribution in the same way we studied \mjpp.
The \mkp\ combinatorial-subtracted distribution (Fig.~\ref{fig:mkp})
shows no significant structure, and it is therefore used to set a limit.
We fit the background from other 
\btojpsikpp\ decays with an exponential function of \mkp\ and 
obtain $N_{peak}=2.9\pm 1.4$. 
The frequentist approach described above, 
with $N_{data}=10$, $N_{comb}=7.8\pm 2.8$ and
$\epsilon / \epsilon_{\psitwos} =1.00 \pm 0.07$
yields the 90\% C.L. limit $\hbox{\ensuremath{\BR(\btojdp)<\brjdp}}$.
This upper limit rules out the explanation of the inclusive 
\jpsi momentum spectrum with intrinsic charm proposed in~\cite{jpsidtheo}.

In summary, we measured 
$\BR(\btojpsikpp) = \brjpsikpp\ $ 
with an error 
almost a factor two smaller than the present average~\cite{PDG} 
and we confirmed the 
observation of $\btoxk$~\cite{belle,cdf}. 
We measured 
$\BR(\btoxk)\times \BR(X\rm{(3872)}\rightarrow \jpsi \pi^+ \pi^-)=\brxk$ 
and $m_{X(3872)}=3873.4 \pm 1.4 \mevcc $.
We also studied the \mjpp\ and \mkp\ distributions searching 
for \btohck\ and \btojdp\ decays and set limits on their \brf s,
$\BR(\btohck)\times \BR(h_c\rightarrow \jpsi \pip\pim) <\brhck$ and  
$\BR(\btojdp) <\brjdp$ at 90\% C.L.

We are grateful for the excellent luminosity and machine conditions
provided by our \pep2\ colleagues, 
and for the substantial dedicated effort from
the computing organizations that support \babar.
The collaborating institutions wish to thank 
SLAC for its support and kind hospitality. 
This work is supported by
DOE
and NSF (USA),
NSERC (Canada),
IHEP (China),
CEA and
CNRS-IN2P3
(France),
BMBF and DFG
(Germany),
INFN (Italy),
FOM (The Netherlands),
NFR (Norway),
MIST (Russia), and
PPARC (United Kingdom). 
Individuals have received support from CONACyT (Mexico), A.~P.~Sloan Foundation, 
Research Corporation,
and Alexander von Humboldt Foundation.


\begin{thebibliography}{99}
\bibitem{cc}Charge-conjugate reactions are included implicitely throughout 
this Letter.
\bibitem{belle} Belle Collaboration, S.~K.~Choi $et\ al.$,   
\myprl\ 91 262001 2003 .
\bibitem{cdf} CDF Collaboration, D.~Acosta $et\ al.$, 
hep-ex/0312021, submitted to Phys. Rev. Lett. 

\bibitem{quigg} E.~J.~Eichten, K.~Lane, and C.~Quigg, 
Phys. Rev. Lett. \textbf{89}, 162002 (2002);
 E.~J.~Eichten, K.~Lane, and C.~Quigg, hep-ph/0401210.
\bibitem{pakvasa} 
S.~Pakvasa and M.~Suzuki, Phys.Lett. B \textbf{579}, 67 (2004);
N.~A.~Tornqvist,  hep-ph/0308277; 
F.~E.~Close and P.~R.~Page, Phys.Lett. B \textbf{578}, 119 (2004);
E.~S.~Swanson, hep-ph/0311229; 
T.~Barnes and S.~Godfrey, hep-ph/0311162;
M.~B.~Voloshin, Phys. Lett. B \textbf{579}, 316 (2004).

\bibitem{hc}
T.~A.~Armstrong {\it et al.},
\myprl  69 2337 1992 .
\bibitem{jpsiincl}
$\babar$ Collaboration, B.~Aubert {\it et al.}, 
\myprd 67 032002 2003 ;
CLEO Collaboration, R.~Balest {\it et al.},
\myprd  52 2661 1995 .
\bibitem{jpsidtheo}
C.~H.~Chang and W.~S.~Hou,
\myprd 64 071501 2001 .
\bibitem{babar-det}$\babar$ Collaboration, B. Aubert $et\ al.$, Nucl. Instrum. \& Methods A \textbf{479}, 1 (2002).
\bibitem{geant}
GEANT4 Collaboration, S.~Agostinelli {\it et al.},
Nucl. Instrum. \& Methods A \textbf{506}, 250 (2003).
\bibitem{babar-charmonium}
$\babar$ Collaboration, B. Aubert $et\ al.$, Phys. Rev. D \textbf{65}, 032001 (2002).
\bibitem{PDG}
Particle Data Group, K.\ Hagiwara {\em et al.}, Phys. Rev. D {\bf 66}, 010001 (2002),
as updated for 2003 at http://pdg.lbl.gov/
\bibitem{PDGstat}
Particle Data Group, K.\ Hagiwara {\em et al.}, Phys. Rev. D {\bf 66}, 010001 (2002), section 31.4.2.
\bibitem {cry}
Crystal Ball Collaboration, T.~Skwarnicki,
 DESY F31-86-02.       

\bibitem{Argus}
ARGUS Collaboration, H.~Albrecht {\em et al.}, \zp C {\bf 48}, 543 (1990).


\end{thebibliography}
\end{document}